\documentclass[useAMS,usenatbib]{mn2e}
\usepackage{times,aas_macros}
\usepackage{rotating}
\input{epsf}

\title[]
{Confirmation of the nature of the absorber in IRAS 09104+4109}

\author[C.-Y. Chiang et al.]
{Chia-Ying Chiang$^{1}\thanks{E-mail: cychiang@ast.cam.ac.uk}$, E. M. Cackett$^{2}$, P. Gandhi$^{3}$, A. C. Fabian$^{1}$\\
$^{1}$Institute of Astronomy, University of Cambridge, Madingley Road, Cambridge CB3 0HA\\
$^{2}$Department of Physics and Astronomy, Wayne State University, 666 W. Hancock St, Detroit, MI 48201, USA\\
$^{3}$Institute of Space and Astronautical Science, Japan Aerospace Exploration Agency, 3-1-1 Yoshinodai, chuo-ku, Sagamihara, Kanagawa 252-5210, Japan\\
}

\date{Accepted 2013 January 14.  Received 2013 January 10; in original form 2012 November 22}
\pagerange{\pageref{firstpage}--\pageref{lastpage}} \pubyear{2011}

\begin{document}

\topmargin = -0.5cm

\maketitle

\label{firstpage}

\begin{abstract}
We present the first long \emph{Suzaku} observation of the
hyperluminous infrared galaxy IRAS 09104+4109 which is dominated by
a Type 2 AGN. The infrared to X-ray SED indicates that the source is
an obscured quasar with a Compton-thin absorber. However, the
3$\sigma$ hard X-ray detection of the source with the
\emph{BeppoSAX} PDS suggested a reflection-dominated, Compton-thick
view. The high-energy detection was later found to be possibly
contaminated by another Type 2 AGN, NGC 2785, which is only 17
arcmin away. Our new \emph{Suzaku} observation offers simultaneous
soft and hard X-ray coverage and excludes contamination from NGC
2785. We find that the hard X-ray component is not detected by the
\emph{Suzaku} HXD/PIN (effective energy band 14-45 keV). Both
reflection and transmission models have been tested on
the latest \emph{Suzaku} and \emph{Chandra} data. The 0.5-10 keV
spectrum can be well modelled by the two scenarios. In addition, our
analysis implied that the absorption column required in both models
is $N_{\rm H} \sim 5 \times 10^{23}$ cm$^{-2}$. Unless IRAS
09104+4109 is a ``changing-look" quasar, we confirm that it is a
Compton-thin AGN. Although the lack of detection of X-ray emission
above 10 keV seems to favour the transmission scenario, we found
that the two models offer fairly similar flux predictions over the
X-ray band below $\sim$ 40 keV. We also found that the strong iron
line shown in the \emph{Suzaku} spectrum is in fact a blend of two
emission lines, in which the 6.4 keV one is mostly contributed from
the AGN and the 6.7 keV from the hot cluster gas. This implies that
the neutral line is perhaps caused by disc reflection, and the
reflection-dominated model is more likely the explanation. The
transmission model should not be completely ruled out, but a deeper
hard X-ray spectrum observation is needed to discriminate between
the two scenarios.

\end{abstract}


\section{Introduction}

The hyperluminous infrared (IR) galaxy IRAS 09104+4109 (z = 0.442)
hosted in a cD galaxy in a rich cluster \citep{Kleinmann88} is one
of the most powerful (IR luminosity $> 10^{46}$ erg s$^{-1}$)
sources within z = 0.5. The active nucleus of IRAS 09104+4109 was
classified as an obscured Type 2 quasar based on its optical
spectrum characteristic \citep{Kleinmann88,HW93,Tran00}, while the
near IR spectrum gave a consistent classification \citep{Soifer96}.
The \emph{ASCA} observation showed that the source is bright in
X-rays, with a powerlaw spectrum which is typically seen in most
AGNs \citep{Fabian94}. \citet{HW93} analysed the VLA 1.4
and 5 GHz observations and found a double-lobed radio source with
straight jets extending along north-west and south-east of the
nucleus. The \emph{ROSAT} High Resolution Imager (HRI) observation
revealed spatial extended X-ray emission with a central dip in the
X-ray intensity \citep{FC95}, which has been confirmed to be one of
two cavities in north-west and south-east of the core with a new 76
ks \emph{Chandra} observation \citep{HL12}. \citet{O'Sullivan12}
presented new GMRT observations and claimed that these X-ray
cavities coincide with the radio jets, while \citet{HL12} indicated
that the cavities are coincident with hotspots. The peaked X-ray
surface brightness profile resolved by \emph{ROSAT} implied that the
cluster has a strong cool core \citep{FC95,Crawford96}.


A strong emission line, probably associated with the
reflection from helium-like iron, was discovered in the \emph{ASCA}
spectrum. The similar fluxes found in \emph{ASCA} (extrapolated) and
\emph{ROSAT} observations imply that the 0.1-10.0 keV X-ray emission
is also likely contributed by the hot gas in the cluster, though the
strong iron line might come from a hidden nucleus.
\citet{Franceschini00} reported a 3$\sigma$ detection of a
transmitted component above 10 keV by the \emph{BeppoSAX} PDS, which
indicated that the spectrum should be interpreted as a reflection
continuum absorbed by a Compton-thick absorber. The equivalent width
of the strong iron line given by the 9.1 ks \emph{Chandra}
observation \citep{Iwasawa01} is $\sim$ 1-2 keV, which further
supports the reflection-dominated scenario. However, the limited
0.1-10.0 keV X-ray spectra are not sufficient in distinguishing the
transmission- and the reflection-dominated models.
\citet{Piconcelli07} analysed the 14 ks \emph{XMM-Newton}
observation and found no significant preference to these models.
They indicated a possibility of ``contamination" from nearby
sources, as the field of view of the PDS instrument of
\emph{BeppoSAX} is large (1 degree $\times$ 1 degree). Also, the
2-10 keV luminosity measured by the transmission-dominated model is
consistent with the expected value on the basis of the bolometric
luminosity. The absorber along the line of sight of the nucleus of
IRAS 09104+4109 is either Compton-thin, or changed from
Compton-thick to -thin between the \emph{BeppoSAX} and
\emph{XMM-Newton} observations in five years.


\citet{Vignali11} found that both \emph{Chandra} and optical/mid-IR
spectral analysis imply heavy, but not Compton-thick obscuration
(i.e. that $N_{\rm H}> 10^{24}$ cm$^{-2}$). Moreover, they presented
a 54-month \emph{Swift} BAT map (15.0-30.0 keV band) which shows
that the hard X-ray emission detected by \emph{BeppoSAX} is probably
associated with a z = 0.009 Type 2 AGN NGC 2785, which is only 17
arcmin away from IRAS 09104+4109. In order to investigate the
ambiguous detection by \emph{BeppoSAX} above 10 keV in depth, data
from the hard X-ray band, where \emph{XMM-Newton} and \emph{Chandra}
have no coverage, are necessary.

The recent observation with \emph{Suzaku} offers simultaneous X-ray
data of 0.5-10.0 keV and 14.0-45.0 keV of the source. Given that
the field of views of \emph{Suzaku} X-ray Imaging Spectrometer (XIS)
and PIN are 17.8 arcmin $\times$ 17.8 arcmin and 34 arcmin
$\times$ 34 arcmin, respectively, emission from the nearby AGN NGC
2785 is likely included in the PIN detector if the pointing is aimed
at IRAS 09104+4109 (RA DEC = 09 13 45.49  40 56 28.2). The
\emph{Suzaku} PIN is a non-imaging instrument and emission from
other sources cannot be simply excluded by region extraction.
Therefore we shifted the pointing 6 arcmin west (new RA DEC
= 09 13 21.49  40 56 28.2) and assigned a $90^{\circ}$ roll angle in
our observation. The entire source still lies within the XIS
detector after the change of pointing, and contribution from NGC
2785 should be excluded from the field of view of PIN.

Once the hard X-ray detection of IRAS 09104+4109 by the
\emph{BeppoSAX} PDS is also seen by the \emph{Suzaku} PIN, it would
confirm the reflection-dominated picture. Compton-thick AGN are
important in understanding the evolution of suppermassive black
holes. So far these objects are mostly found at low redshifts
\citep[z $<$ 0.05; ][]{Comastri04}. If IRAS 09104+4109 is
Compton-thick, it will be one of the few distant Compton-thick AGN
to have reliable X-ray spectroscopic data. Relatively-few such
objects are known at the moment. Luminous Compton-thick AGN with
well-studied multi-wavelength data are useful for understanding an
important evolutionary phase in the growth of supermassive black
holes.

This paper presents analysis for the first \emph{Suzaku} long
observation of IRAS 09104+4109 and offers conclusive results about
the nature of the absorber in the source. It has been debated for
years that the X-ray spectrum is transmission- or
reflection-dominated, and we will discuss this in detail in section
\ref{discussion}. Calculations in this paper were assumed a flat
cosmology with $H_{0}$ = 71 km s$^{-1}$ Mpc$^{-1}$,
$\Omega_{\lambda} = 0.73$ and $\Omega_{0} = 0.27$.

\section{Data Reduction} \label{section_da}

\subsection{\emph{Suzaku}}

IRAS 09104+4109 was observed with \emph{Suzaku} during 2011 November
18-20 resulting in a total of $\sim$113 ks of good exposure time.
The XIS was operated in the normal mode, and both $3\times3$ and
$5\times5$ editing modes were operated in all detectors (XIS0, XIS1
and XIS3). As mentioned in the previous section, the pointing has
been shifted by 6 arcmin. The source still lies perfectly within the
XIS field of view, and no other point sources can be seen.
Data have been reduced using the {\sevensize HEASOFT V6.11.1}
software package with the latest calibration database following the
standard procedure.
The source spectrum was extracted using {\sevensize XSELECT} from a
circular region with a radius of 110 arcsec. A larger circular
region with a 180-arcsec radius has been used for background
estimation from a source-free region. Response files were produced
by the script {\sevensize XISRESP}, which calls {\sevensize
XISRMFGEN} and {\sevensize XISSIMARFGEN} to generate redistribution
matrix files (RMF) and ancillary response files (ARF), respectively.
We combined spectra and response files of the two front-illuminated
(FI) CCD XIS detectors (XIS0 and XIS3) using the script {\sevensize
ADDASCASPEC} in {\sevensize FTOOL}. The spectral bins of all spectra
were grouped for at least 20 counts. We use the FI spectrum over
0.5-10.0 keV, and the BI spectrum over 0.5-7.0 keV for further
analysis, as the signal-to-noise ratio of the BI data is lower at
high energies.


\begin{figure}
\begin{center}
\leavevmode \epsfxsize=8.5cm \epsfbox{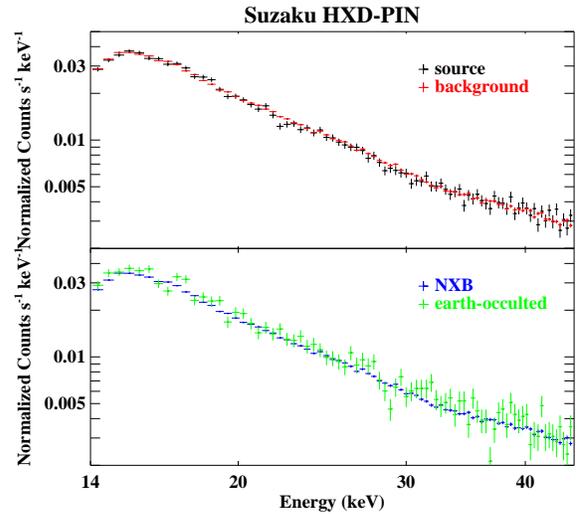}
\end{center}
\caption{The upper panel shows the PIN data (before
background  correction) in black points and the total background in
red points. The lower panel compares the Non-X-ray background and
the earth-occulted background, which are shown in blue and green
points, respectively.} \label{pin}
\end{figure}

The Hard X-ray Detector (HXD) was operated in XIS-nominal pointing
mode. A non-X-ray background (NXB) and a cosmic X-ray background
(CXB) should be combined to form the total HXD/PIN background
spectrum. We obtained the NXB event file directly from the
\emph{Suzaku} Data Centre and extracted it using the background
model D (the tuned model). The CXB was simulated using the PIN
response for flat emission distribution. The count rate of source
before background correction is $0.340 \pm 0.002$ counts s$^{-1}$ in
the effective PIN energy band 14.0-45.0 keV (all the count rates of
PIN will be given over this band if not specified), appearing below
the total background level ($0.343 \pm 0.001$ counts s$^{-1}$). As
the count rates of the source and the background are at similar
level (see also the upper panel in Fig. \ref{pin}), we produced the
earth-occulted background for a further check. The earth-occulted
background is ideally consistent with the NXB, and it should replace
the NXB if discrepant. The lower panel in Fig. \ref{pin} shows that
the earth-occulted background (green data points) and the NXB (blue
points) are in good agreement. The count rate of the earth-occulted
background ($0.344 \pm 0.004$ counts s$^{-1}$) is slightly higher
than that of the NXB ($0.325 \pm 0.001$ counts s$^{-1}$), confirming
the non-detection in the PIN band. Thus we do not include the PIN
spectrum in all our spectral fittings. Although the hard X-ray
emission from IRAS 09104+4109 is beyond detection in the HXD/PIN, we
can estimate an upper limit of flux by including the systematic
error of the NXB and CXB. We generated an ARF for the HXD/PIN using
the {\sevensize HXDARFGEN} tool. The output ARF was coupled with the
"hxdnominal" response file "ae\_hxd\_pinhxnome11\_20110601.rsp" to
account for the off-axis correction. The resulting 14.0-45.0 keV
fluxes of the NXB and CXB are around $2.16 \times 10^{-10}$ ergs
cm$^{-2}$ s$^{-1}$ and $9.23 \times 10^{-12}$ ergs cm$^{-2}$
s$^{-1}$. Considering the systematic error of the NXB to be 3\% (for
observations with more than 10 ks exposure time
\footnote{http://heasarc.gsfc.nasa.gov/docs/suzaku/prop\_tools/suzaku\_td/})
and the lower limit of the total background flux can be calculated.
The background-uncorrected 14.0-45.0 keV flux of the source is $\sim
2.23 \times 10^{-10}$ ergs cm$^{-2}$ s$^{-1}$, and we hence estimate
the upper limit of the background-corrected source flux to be $\sim
4.25 \times 10^{-12}$ ergs cm$^{-2}$ s$^{-1}$. The observed
\emph{BeppoSAX} PDS 20-100 keV flux is $\simeq 10^{-11}$ ergs
cm$^{-2}$ s$^{-1}$, which is higher than the PIN upper limit $\sim
6.26 \times 10^{-12}$ ergs cm$^{-2}$ s$^{-1}$ in the same band.

\subsection{\emph{Chandra}}

The \emph{Chandra} Advanced CCD Imaging Spectrometer (ACIS) observed
IRAS 09104+4109 on 2009 January 6 for about 76 ks in very faint
mode. We used tools in the latest {\sevensize CIAO V4.4} software
package to reduce the data. The observation ends up with a good
exposure of $\sim$ 72 ks after background flare filtering. The
nucleus of the cluster appears as a point source in the 0.5-7.0 keV
(effective energy band of the ACIS) image. Since IRAS 09104+4109 is
a strong cool-core cluster showing a significant radial temperature
gradient, background subtraction could be critical in some
circumstances. Thank to the excellent point spread function (PSF) of
the \emph{Chandra} ACIS, the nucleus spectrum can be extracted from
a circular region with a 1-arcsec radius. In this case, the
background was extracted from the surrounding 2-4 arcsec annulus in
order to avoid possible source emission outside the 1-arcsec-radius
region and collect enough counts for the chi-statistics. The
resulting spectrum contains emission from the nucleus of the
cluster, and has excluded the thermal contribution from nearby hot
gas. We also generated a spectrum with a 55-arcsec radius source
region and a 55-110 arcsec annulus background region for comparison
to our \emph{Suzaku} data (regions used in the \emph{Suzaku}
observation are too large and would include edges of the chips of
the ACIS). All the point sources resolved by \emph{Chandra} within
the background region have been removed for clean background
extraction. Finally, we grouped both of our \emph{Chandra} spectra
with a minimum of 20 counts per bin as we did for the \emph{Suzaku}
data.

\begin{figure}
\begin{center}
\leavevmode \epsfxsize=8.5cm \epsfbox{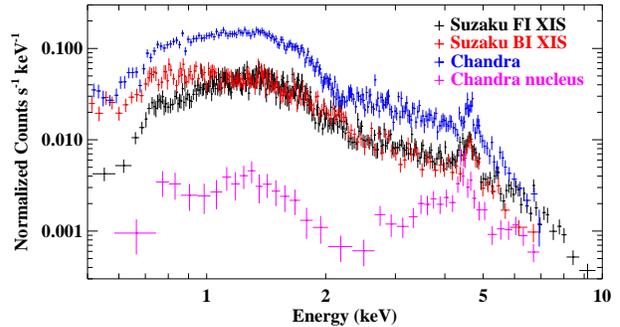}
\end{center}
\caption{The figure shows the spectra we extracted from both
\emph{Suzaku} and \emph{Chandra} Observations. The data points have
been mildly re-binned for clarity. The differences below $\sim$ 1
keV between the \emph{Suzaku} FI and BI XIS spectra are due to
different calibrations. We ignore the \emph{Suzaku} BI XIS data
above 7 keV due to poor signal-to-noise ratio.} \label{data}
\end{figure}

The \emph{Chandra} spectra obtained by different reduction
strategies are shown in Fig. \ref{data}. There seems to be a drop
between the $<$ 2 keV and the $>$ 3 keV band in the nucleus
spectrum, but it is not obvious in the spectrum extracted from the
55-arcsec radius region (hereafter the \emph{Chandra}
55-arcsec spectrum). We also plotted both the \emph{Suzaku} FI XIS
and back-illuminated (BI) XIS spectra in Fig. \ref{data}. The shape
of the \emph{Suzaku} FI XIS spectrum is fairly similar to that of
the \emph{Chandra} 55-arcsec spectrum, which is expected because
they were both extracted from a region including the nucleus and
nearby hot gas. The flux difference between them is likely due to
different background subtraction areas. A clear red-shifted iron
line appears in all spectra. Interestingly, the iron line in the
\emph{Chandra} nucleus spectrum seems to peak at a different energy
from that of the 55-arcsec and \emph{Suzaku} spectra. This will be
further investigated in the following section.

\begin{figure}
\begin{center}
\leavevmode \epsfxsize=8.5cm \epsfbox{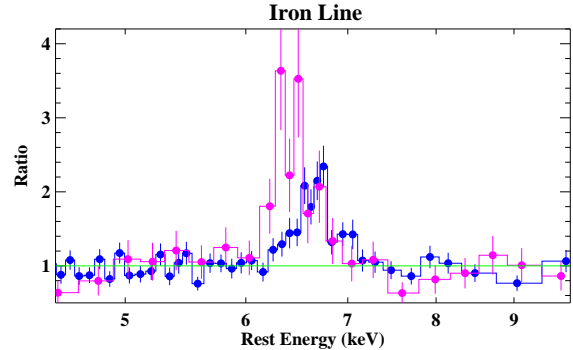}
\end{center}
\caption{The figure shows the data/model ratio of the \emph{Chandra}
nucleus (magenta points) and 55-arcsec (blue points) spectra, which
have been fitted with a simple powerlaw. It can be seen that the
line energies are different.
              } \label{line}
\end{figure}

\begin{table*}
  \caption{The table lists fitting parameters that are bound in all
  data. $N_{\rm H}$ in this table shows the absorption column of
  {\sevensize ZPHABS}. $\Gamma$ is the photon index of the powerlaw
  component. $kT_{1}$ and $kT_{2}$ are the temperatures of the two
  thermal plasma components. Errors have been calculated at the 90
  per cent confidence level.
  }
  \label{fitting_para}
  \begin{tabular}{cccrccccc}
  \hline
  \hline

Model & $N_{\rm H} (10^{22}$ cm$^{-2})$  & $\Gamma$ &  $kT_{1}$ (keV) & $kT_{2}$ (keV)  & $Z/Z_{\odot}$ & $\chi^{2}/d.o.f.$  \\
\hline Reflection & $45.3^{+22.9}_{-12.0}$  & $1.59^{+0.31}_{-0.19}$
&
$2.22^{+1.16}_{-0.43}$ & $6.36^{+1.64}_{-0.90}$ & $0.42^{+0.07}_{-0.06}$ &  1065.5/1152  \\
Transmission & $53.5^{+23.0}_{-21.6}$  & $1.28^{+0.63}_{-0.65}$ &
$3.25^{+1.18}_{-1.23}$ & $7.03^{+2.97}_{-1.67}$ & $0.68^{+0.17}_{-0.12}$ & 1072.4/1150   \\
\hline \hline
\end{tabular}
\end{table*}

\section{Data Analysis and Comparison}

\subsection{Iron Line}

A strong iron line with an equivalent width $EW \sim$ 300-1200 eV
\citep[e.g.][]{Franceschini00,Iwasawa01,Piconcelli07} has been
reported in all previous X-ray studies of this source. The
equivalent width is a model-dependent quantity, causing differences
in values quoted in previous literature. We first fitted a simple
model composed of a powerlaw and a Gaussian line to the 3.0-7.0 keV
band of our data. The rest line energy of the \emph{Suzaku} spectra
is $6.63^{+0.03}_{-0.04}$ keV, consistent with the value
$6.64\pm0.05$ keV given by the \emph{Chandra} 55-arcsec spectrum.
Nevertheless, the rest line energy of the iron line in the
\emph{Chandra} nucleus spectrum is $6.42\pm0.06$ keV instead of
$\sim$ 6.6 keV found in other spectra. We plot the data/model
(fitted with a powerlaw only) ratio of the \emph{Chandra} data in
Fig. \ref{line}, in which the different peaks of the line can be
seen. As the \emph{Chandra} 55-arcsec and \emph{Suzaku} spectra also
include emission from the nucleus, the iron line shown in these
spectra should be formed by two components, in which one of them is
attributed to the nucleus and the other to the diffuse hot gas. In
most literature to date, the iron line shown in the X-ray spectra of
IRAS 09104+4109 has been treated as a single line instead of two. It
is not surprising that the line energy and equivalent width of the
iron line obtained from different data sets are in disagreement.
\citet{Vignali11} also suggested the possibility that the iron
emission line is originated by a blend of emission features from
both the central AGN and cluster gas.

We hereafter decomposed the iron line shown in the \emph{Chandra}
55-arcsec spectrum into two Gaussian components and re-fitted the
data. The rest line energy of the second Gaussian component turned
out to be $6.70^{+0.06}_{-0.10}$ keV. The equivalent widths obtained
for both iron line components are $EW = 82^{+50}_{-37}$ eV for the
6.42 keV line, and $EW = 227^{+173}_{-81}$ eV for the 6.70 keV line,
respectively. As for \emph{Suzaku} data, the equivalent widths are
$EW = 87^{+45}_{-39}$ eV for the 6.42 keV line and $EW =
260^{+115}_{-99}$ eV for the 6.70 keV line, which are consistent
with the \emph{Chandra} 55-arcsec spectrum. The equivalent
width found in the \emph{Chandra} nucleus spectrum is high ($EW =
338^{+152}_{-133}$ eV) possibly owing to a lower continuum level.
Although in Fig. \ref{line} it seems that the lines in both spectra
are of similar line widths, the line shown in the nucleus spectrum
can be fitted by a narrow (line width $<$ 0.1 keV) Gaussian line and
no further components are required. In the following analysis, we
model the iron line in the 55-arcsec and \emph{Suzaku} spectra with
two decomposed line components.

\subsection{Spectral Fitting} \label{spec_fitting}

We started detailed spectral analysis by constructing a model to
explain the \emph{Chandra} nucleus spectrum. The origin of the 6.42
keV Fe K line is generally attributed to reflected emission from
cold matter illuminated by high energy photons. Hence a neutral
reflection continuum might be needed in the model. In addition, the
structures in the hard ($>$ 3 keV) and soft ($<$ 2 keV) energy band
in the \emph{Chandra} nucleus spectrum shown in Fig. \ref{data}
seems to be symmetric, implying that the $<$ 2keV component
can be the red-shifted Fe L complex caused by strong reflection.
\citet{Iwasawa01} also suggested this possibility in their analysis
of the $\sim$ 9 ks \emph{Chandra} observation. We fit a model
composed of an absorbed powerlaw component {\sevensize POWERLAW} and
a reflection continuum {\sevensize REFLIONX} \citep{Ross05} in
{\sevensize XSPEC} to the \emph{Chandra} nucleus spectrum. The
photon index in the {\sevensize REFLIONX} component has been set
identical with that in the {\sevensize POWERLAW} component, and
allowed to vary in a range between 1.4 and 3.3 which is reasonable
for the reflection scenario. The ionisation parameter $\xi$ in
{\sevensize REFLIONX} was set to unity ($\xi$ = 1.0) to model the
6.4 keV Fe-K emission line from a neutral reflector with solar
abundance. The Galactic absorption column $N_{\rm H} = 1.8 \times
10^{20}$ cm$^{-2}$ \citep{Murphy96} is also included and modelled by
{\sevensize
TBNEW}\footnote{http://pulsar.sternwarte.uni-erlangen.de/wilms/research/tbabs/}
(Wilms, Juett, Schulz, Nowak, 2012, in preparation) with the Wilms
abundance. The absorption component in the source has been modelled
using {\sevensize ZPHABS}. Nevertheless, the model cannot, even with
a high iron abundance, explain the low-energy excess in the spectrum
well. It seems more likely that the low-energy hump in the core-only
spectrum is caused by photonionised gas. We added a thermal
plasma component {\sevensize MEKAL} \citep[][ see also Kaastra
1992\footnote{http://www.sron.nl/files/HEA/SPEX/physics/megron.pdf}]
{Mewe85,Liedahl95} which includes line emission from several
elements and the Fe-L complex to the model and obtained a better
fit.

\begin{table}
\caption{The table below lists the fitting parameters which have not
been shown in Table \ref{fitting_para}. In this table $N_{\rm pow}$,
$N_{\rm mek1}$, $N_{\rm mek2}$, and $N_{\rm ref}$ represent the
normalisation of the powerlaw component, the two {\sevensize MEKAL}
components and the reflection component, respectively. $F_{\rm 14-45}$
and $F_{\rm 20-80}$ are the 14-45 keV and 20-80 keV fluxes predicted
by the model. All fluxes are shown in ergs cm$^{-2}$ s$^{-1}$.
* Since the predicted values for \emph{Suzaku} XIS FI and BI spectra
are fairly similar, we quoted that of the FI spectrum only.
 }
\label{norm}
\begin{tabular}{@{}lccc}
\hline\hline
Parameter & \emph{Suzaku} & \emph{Chandra} & \emph{Chandra} \\
 & FI \& BI & 55-arcsec & nucleus\\
\hline
 & \multicolumn{3}{c}{Reflection}\\
$N_{\rm pow}$   & \multicolumn{2}{c}{$7.9^{+10.0}_{-7.0}\times 10^{-5}$} & $2.4^{+3.2}_{-1.3}\times 10^{-4}$ \\
$N_{\rm mek1}$ & \multicolumn{2}{c}{$8.1^{+9.9}_{-4.1}\times 10^{-4}$}  & $1.3^{+0.2}_{-0.3}\times 10^{-4}$\\
$N_{\rm mek2}$ & \multicolumn{2}{c}{$2.7^{+0.4}_{-1.0}\times 10^{-3}$}  & 0 (fixed) \\
$N_{\rm ref}$     & \multicolumn{2}{c}{$3.3^{+2.0}_{-1.1}\times 10^{-5}$}  & $3.7^{+3.1}_{-1.7}\times 10^{-5}$ \\
$F_{\rm 14-45}$ & *$2.37 \times 10^{-12}$ & $2.10 \times 10^{-12}$ & $3.20 \times 10^{-12}$ \\
$F_{\rm 20-80}$ & *$2.89 \times 10^{-12}$ & $2.55 \times 10^{-12}$ & $4.18 \times 10^{-12}$ \\
\hline
 & \multicolumn{3}{c}{Transmission}\\
$N_{\rm pow}$   & \multicolumn{2}{c}{$9.5^{+27.5}_{-7.4}\times 10^{-5}$} & $1.6^{+4.3}_{-1.1}\times 10^{-4}$ \\
$N_{\rm mek1}$ & \multicolumn{2}{c}{$1.5^{+1.6}_{-1.0}\times 10^{-3}$}  & $(1.4\pm0.2)\times 10^{-4}$\\
$N_{\rm mek2}$ & \multicolumn{2}{c}{$2.0^{+1.0}_{-1.1}\times 10^{-3}$}  & 0 (fixed) \\
$F_{\rm 14-45}$ & *$1.89 \times 10^{-12}$ & $1.68 \times 10^{-12}$ & $2.78 \times 10^{-12}$ \\
$F_{\rm 20-80}$ & *$3.12 \times 10^{-12}$ & $2.76 \times 10^{-12}$ & $4.70 \times 10^{-12}$ \\

\hline\hline
\end{tabular}
\end{table}

\begin{figure*}
    \begin{center}
        \begin{minipage}{1\textwidth}
            \begin{minipage}{0.5\textwidth}
                \begin{center}
                    \leavevmode \epsfxsize=8.5cm \epsfbox{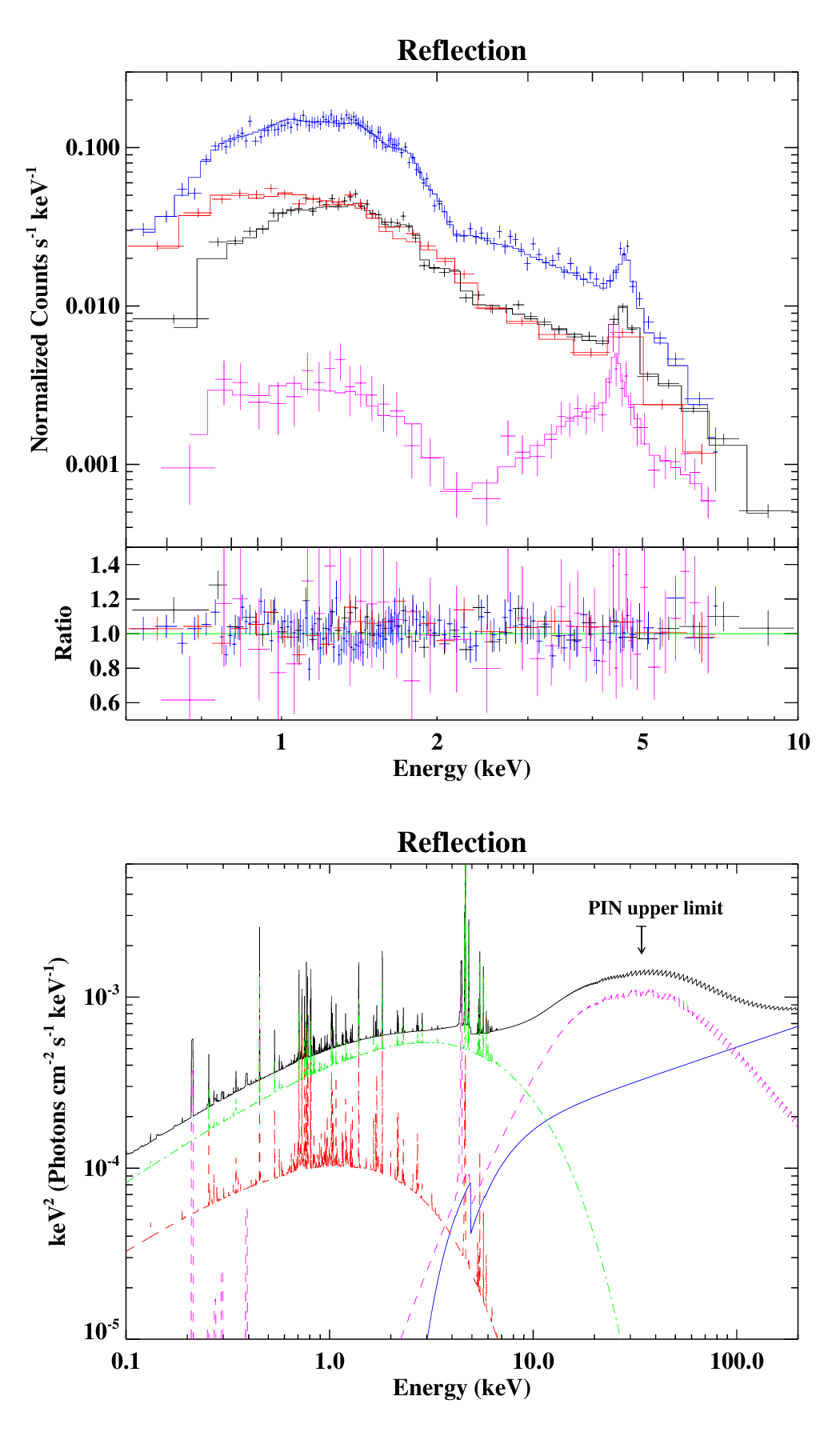}
                \end{center}
            \end{minipage}
            \begin{minipage}{0.5\textwidth}
                \begin{center}
                    \leavevmode \epsfxsize=8.5cm \epsfbox{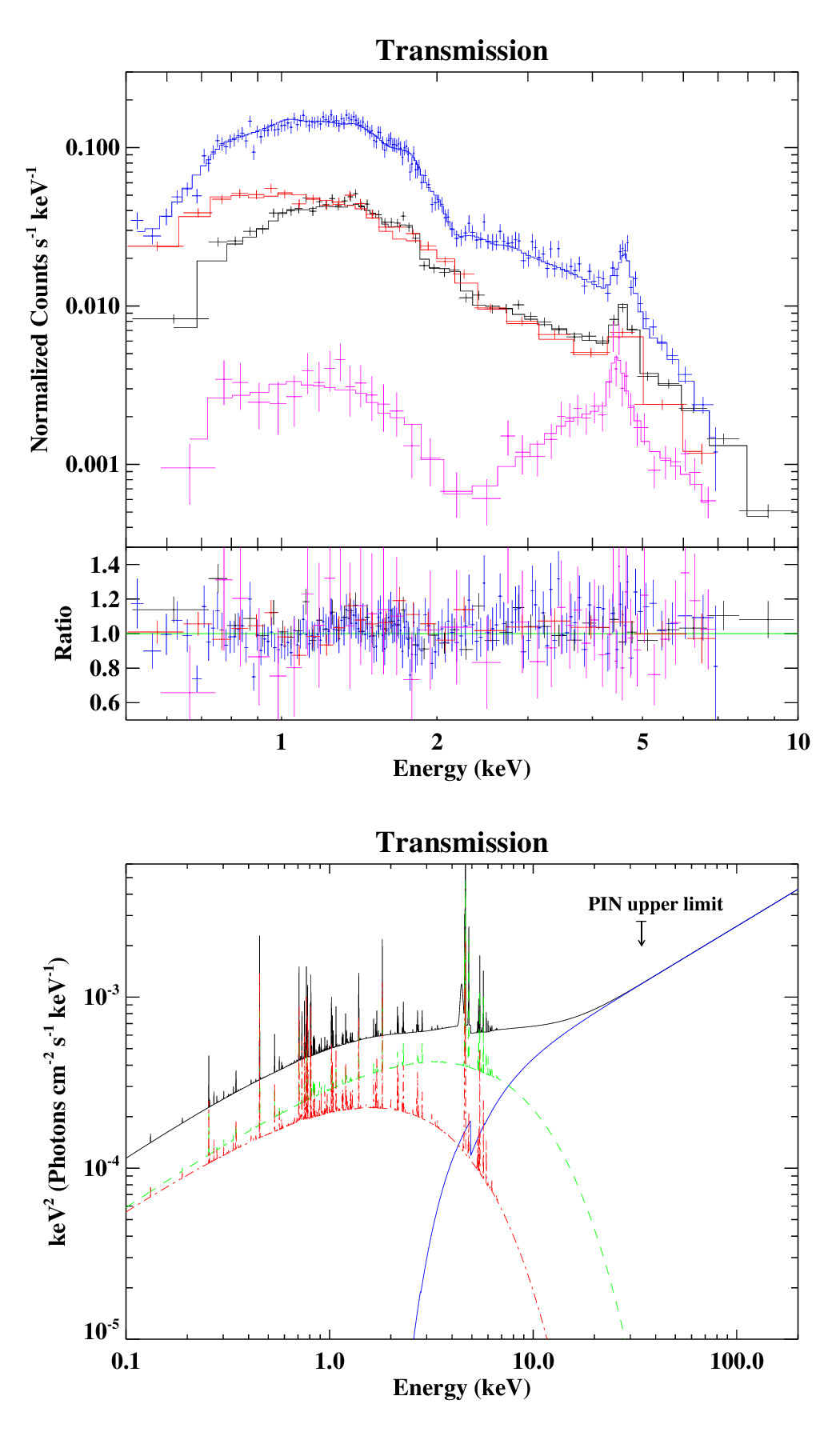}
                \end{center}
            \end{minipage}
            \caption{The upper set of figures shows the spectral fittings of
            the reflection and transmission models, while the lower set shows
            the decomposed components of the models. In the two upper plots,
            colours of each spectrum are expressed in the same way as in Fig.
            \ref{data}. In both lower figures, the black solid line stands the
            total model and coloured ones represent contributions
                          from different components. The green and red dot-dash lines show the two {\sevensize MEKAL}
                          components of different temperatures. We plot the absorbed powerlaw as a blue solid line in both
                          panels, and the reflection component has been shown as magenta dash line. The Galactic absorption
                          has been taken out in these figures.}
            \label{model}
        \end{minipage}
    \end{center}
\end{figure*}

In order to extend the model to a version that can explain the
\emph{Chandra} 55-arcsec and \emph{Suzaku} spectra, the 6.7 keV Fe
XXV emission line shown in these spectra should be modelled. If the
line originates from reflection, it should come from part of a
highly-ionised accretion disc which is usually close to the central
black hole, where the illuminating source is nearby. By comparing
the \emph{Chandra} spectra extracted from different source regions,
it is obvious that the 6.7 keV line is generated from the
environment around the nucleus but not the central AGN. In addition,
the X-ray emitting region of a typical AGN is expected to be small,
and a spectrum extracted from 2-4 arcsec, which corresponds to
$\sim$20-30 kpc, should contain no X-ray emission from the nucleus
AGN. Hence the 6.7 keV is more likely scattered by the hot diffuse
gas instead of reflected from the accretion disc. We then added
another {\sevensize MEKAL} component with a high gas temperature
into the model to fit the 6.7 keV line. The resulting full model can
be expressed as: {\sevensize TBNEW}*( {\sevensize
ZPHABS}*{\sevensize POWERLAW} + {\sevensize REFLIONX} + {\sevensize
MEKAL}$_{1}$ + {\sevensize MEKAL}$_{2}$). We set the normalisation
of {\sevensize MEKAL}$_{2}$ to be 0 for the \emph{Chandra} nucleus
spectrum, as this component is not required here. The \emph{Chandra}
and \emph{Suzaku} observations differ by about two years, which is
short for a quasar to evolve dramatically. We expected the
\emph{Chandra} and \emph{Suzaku} data to be slightly different in
flux only. The main variables in the model, that are, the absorption
column $N_{\rm H}$ of {\sevensize ZPHABS}, the photon index
$\Gamma$, and the temperature $kT$ and the metallicity $Z/Z_{\odot}$
in the {\sevensize MEKAL} component, are bound in all data set. We
assume that there is only an instrumental cross-calibration constant
between the \emph{Chandra} 55-arcsec spectrum and \emph{Suzaku}
data, and bind normalisations of all model components together. As
for the \emph{Chandra} nucleus spectrum, normalisations of the
powerlaw component, the {\sevensize REFLIONX} component, and the
{\sevensize MEKAL}$_{1}$ component are allowed to vary in a way
different from the other spectra. We summarised the fitting
parameters of the reflection model in Table \ref{fitting_para}, and
normalisations of parameters in Table \ref{norm}.

The reflection model we constructed is similar to the ``absorbed
powerlaw + {\sevensize MEKAL} + reflection" model used in
\citet{Franceschini00}. However, in their work an absorption column
greater than $5 \times 10^{24}$ cm$^{-2}$ is required, while in
Table \ref{fitting_para} it can be clearly seen that the absorption
column required here is $N_{\rm H}\sim 4.5\times10^{23}$ cm$^{-2}$,
which is heavy but Compton-thin. The value is consistent with the
number suggested by the transmission scenario ($N_{\rm H}\sim
5\times10^{23}$ cm$^{-2}$, \citealt{Piconcelli07,Vignali11}). Since
the HXD/PIN has no solid detection in the high-energy band, the
transmission model is a plausible explanation as well. We
constructed a transmission model by replacing the {\sevensize
REFLIONX} component with a Gaussian line using {\sevensize ZGAUSS}
in {\sevensize XSPEC}. The constraint on the range of the photon
index has been lifted. Again we bind the main parameters in all data
as we did when fitting with a reflection model. We allow
normalisations of the powerlaw component, the {\sevensize
MEKAL}$_{1}$ component, and the {\sevensize ZGAUSS} component in the
\emph{Chandra} nucleus spectrum to vary independently, while
normalisations of these components in the other three spectra are
bound together. The results have been again listed in Table
\ref{fitting_para} \& \ref{norm}.

The transmission scenario also suggests a Compton-thin absorber with
an absorption column of $\sim 5.4\times10^{23}$ cm$^{-2}$, which is
consistent with that obtained by the reflection model and previous
studies. Most of the resulting fitting parameters of both models are
close and consistent at the 90\% confidence level. The photon index
needed in each model is slightly different perhaps due to different
broadband continuum (see also Fig. \ref{model}). The same reason may
cause the differences in the temperatures of {\sevensize MEKAL}
components. Because the {\sevensize REFLIONX} generates emission
features in the soft X-ray band ($\sim$ 1 keV) and the iron line
band, temperatures required in the {\sevensize MEKAL} components
might be discrepant when the reflection continuum is replaced. The
{\sevensize MEKAL$_{1}$} component is probably contributed by the
thermal emission in the nucleus (the 1-arcsec-radius circular region),
and the {\sevensize MEKAL$_{2}$} originated from the hot gas in the
region outside the nucleus. The gas temperature of
{\sevensize MEKAL$_{1}$} we obtained by each model is below 4 keV,
which is consistent with the results of \citet{O'Sullivan12}. Our
fitting gives a gas temperature of $\sim$ 7 keV for {\sevensize
MEKAL$_{2}$}, which also lies well in the 5-8 keV range indicated by
\citet{O'Sullivan12}. The values suggested by either the reflection
or the transmission model are reasonable. As for the metallicity,
values obtained by our models are close and only in mild disagreement.
The best-fitted
metallicity of the transmission model is closer to the value
reported in \citet{O'Sullivan12}, but that implied by the reflection
model still lies within their 1 $\sigma$ uncertainties.

The reflection and transmission models we tested in this data set
result in comparable reduced $\chi^{2}$ (see Table
\ref{fitting_para}). Even with a better signal-to-noise ratio, X-ray
data cover across the 0.5-10.0 keV band are not sufficient to
distinguish these two interpretations. The lack of high-energy
coverage makes model selection difficult, and the results are of no
significant statistical difference. Fig. \ref{model} shows the
theoretical predictions from both models, and it can be clearly seen
that they are fairly similar below 10 keV and start to divert above
$\sim$ 40 keV. The hard X-ray fluxes, especially the 20-80 keV band,
predicted by these models are fairly close (see Table \ref{norm})
and both below the upper limit estimated in section
\ref{section_da}. As the results of the two models statistically comparable, we
will discuss in the following section the physical possibility of
both models. The absorption-corrected rest frame 2-10 keV
luminosities (calculated from the numbers shown in Table \ref{lum}) of the nucleus are $\sim (1.3-2.3) \times 10^{44}$ erg
s$^{-1}$ and $\sim (1.4-2.2) \times 10^{44}$ erg s$^{-1}$, predicted by
the reflection and transmission models, respectively. These values
are slightly higher than $(1.2-1.3) \times 10^{44}$ erg s$^{-1}$
which was reported in \citet{Vignali11}, but lower than the
predicted value of $\sim$10$^{45}$ erg s$^{-1}$ obtained by the same
authors based upon IR observations from \citet{Lanzuisi09}. Using
instead the mid-IR:X-ray correlation of local Seyferts from
\citet{Gandhi09} results in an upper-limit to the X-ray power of
$\sim 5 \times 10^{45}$ erg s$^{-1}$. This limit is consistent with
the observations but several times higher, because of the fact that
IR data on this source are from the \emph{Spitzer} mission, which
cannot spatially resolve the nucleus emission from stellar activity
in typical ultra- and hyper-luminous infrared galaxies
\citep[see][]{Vignali11,Gandhi09}.

\begin{table}
\caption{The table lists the absorption-uncorrected rest from 2-10 keV flux *$F_{2-10}$, and the intrinsic 2-10 keV absorption-corrected flux $F_{\rm
2-10}$. $F_{\rm thermal}$ and $F_{\rm nucleus}$ are the decomposed components of $F_{\rm
2-10}$ from thermal emission of the cluster gas and the nucleus.
Fluxes are all shown in ergs cm$^{-2}$ s$^{-1}$.
 }
\label{lum}
\begin{tabular}{@{}lccc}
\hline\hline
Flux & \emph{Suzaku} XIS & \emph{Chandra} 55-arcsec & \emph{Chandra} nucleus \\
\hline
 & \multicolumn{3}{c}{Reflection}\\
$F_{2-10}$* & $2.00 \times 10^{-12}$ & $1.76 \times 10^{-12}$ & $1.04 \times 10^{-12}$ \\
$F_{2-10}$ & $1.74 \times 10^{-12}$ & $1.54 \times 10^{-12}$ & $3.53 \times 10^{-13}$ \\
$F_{\rm thermal}$ & $1.53 \times 10^{-12}$ & $1.36 \times 10^{-12}$ & $2.28 \times 10^{-14}$ \\
$F_{\rm nucleus}$ & $2.08 \times 10^{-13}$ & $1.84 \times 10^{-13}$ & $3.31 \times 10^{-13}$ \\
\hline
 & \multicolumn{3}{c}{Transmission}\\
$F_{2-10}$* & $2.14 \times 10^{-12}$ & $1.89 \times 10^{-12}$ & $9.49 \times 10^{-13}$ \\
$F_{2-10}$ & $1.74 \times 10^{-12}$ & $1.54 \times 10^{-12}$ & $3.53 \times 10^{-13}$ \\
$F_{\rm thermal}$ & $1.52 \times 10^{-12}$ & $1.35 \times 10^{-12}$ & $3.77 \times 10^{-14}$ \\
$F_{\rm nucleus}$ & $2.20 \times 10^{-13}$ & $1.95 \times 10^{-13}$ &  $3.15 \times 10^{-13}$\\

\hline\hline
\end{tabular}
\end{table}



\section{Discussion} \label{discussion}
\subsection{Nature of the Absorber}

The main supporting evidence of the Compton-thick interpretation is
the hard X-ray detection by the \emph{BeppoSAX} PDS instrument. In
our latest \emph{Suzaku} observation, the pointing has been shifted
to avoid possible contamination from the nearby Type II AGN NGC
2785. As a result the \emph{Suzaku} HXD/PIN gave no solid detection
but an estimated upper limit which is still lower than the measured
\emph{BeppoSAX} flux, and the marginal 3$\sigma$ \emph{BeppoSAX}
detection is therefore questionable. The non-detection in the hard
X-ray band also implies that the reflection model used in
\citet{Franceschini00} is not necessarily required to explain the
data. The hard X-ray flux obtained by the \emph{BeppoSAX}
observation was so high that the extrapolated transmission model
failed to reach the flux, and the only possibility was to attribute
the high-energy transmitted component to inverse Compton scattering.
In order to explain the broadband X-ray data, the reflection
continuum has to be heavily absorbed by extreme absorption column
across the X-ray band below 10 keV and produce a Compton hump to
interpret the hard X-ray emission. \citet{Piconcelli07} and
\citet{Vignali11} obtained a similar high-energy flux prediction
using a reflection continuum without absorber. They assumed that the
absorber is of the same nature as the reflector, so using a
reflection continuum automatically implies an optically-thick ($\tau
> 1$) absorber. However, considering that the broad iron line
displayed in the \emph{Suzaku} XIS spectrum is a blend of two
emission lines which can be caused by different mechanisms, the
reflection scenario should not be ruled out. The narrow 6.4 keV Fe-K
line could be reflected from the cold, optically-thick accretion
disc in the central AGN. In this case, it is not controversial to
model the data with a reflection continuum, as the optically-thick
reflector could be part of the accretion disc and not necessarily
act as an absorber. The reflection scenario is completely plausible
here.

\begin{figure}
\begin{center}
\leavevmode \epsfxsize=8.5cm \epsfbox{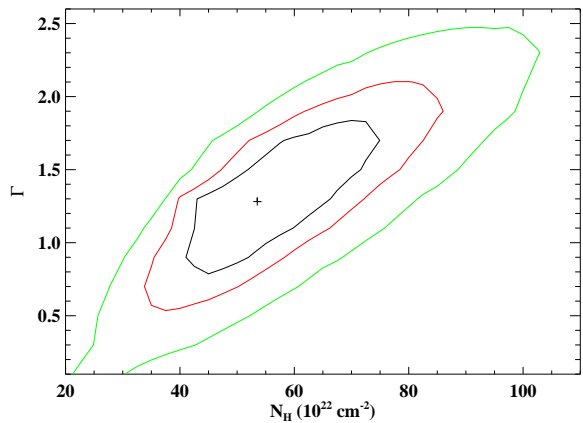}
\end{center}
\caption{The figure shows the contour plot of column density of the
absorber $N_{\rm H}$ against the photon index $\Gamma$. Contours are
plotted at 67 (green), 90 (red) and 99 (black) per cent levels.}
\label{con}
\end{figure}

In our analysis, the absorption columns required in the reflection
and transmission models are not extreme, and both scenarios give
consistent values ($\sim 5 \times 10^{23}$ cm$^{-2}$). The
absorption column is correlated with the photon index, and we plot a
contour of these parameters. It can been seen in Fig. \ref{con} that
the absorption column spans a range of Compton-thin values when the
photon index varies. In previous studies, the reflection model
always links to a Compton-thick absorber, but we showed here that
the reflection scenario does not necessarily imply the Compton-thick
interpretation. As we mentioned before, the reflector needs not to
be the absorber, and it is not surprising that we obtained a
Compton-thin result using the {\sevensize REFLIONX} model. If the
\emph{BeppoSAX} PDS detection is solid, the reflection model is the
only interpretation capable of explaining the high-energy flux.
Nevertheless, as the hard X-ray emission is weak, the reflection
model can still explain the spectrum well. The key to determine the
nature of absorber is the flux above 10 keV, but not the model used
to fit the data. It has been debated for a while that whether the
absorber is Compton-thick or not. Now our analysis gives a
conclusive result that the absorber in IRAS 09104+4109 is
Compton-thin, provided that it is not a ``changing-look" quasar.

\subsection{Reflection or Transmission above 10 keV?}

The two models we tested in this work gave statistically comparable
results and similar fitting parameters. The predicted high-energy
fluxes are both below the upper limit given by the HXD/PIN. We found
that our predictions are also under the upper limit implied by the
\emph{Swift} BAT map \citep{Vignali11}. The 3$\sigma$ upper limit in
the 15-30 keV band implied by the BAT map is $\sim$ 1.9$\times
10^{-12}$ erg cm$^{-2}$ s$^{-1}$, and the predicted fluxes in this
band are (1.2-1.8)$\times 10^{-12}$ and (8.8-14.3)$\times 10^{-13}$
erg cm$^{-2}$ s$^{-1}$ for the reflection and transmission models,
respectively. The reflection scenario has been ruled out in some
studies due to its high flux prediction across the hard X-ray band.
In our study, the reflection model works as well as the transmission
model while no hard X-ray detection has been confirmed. The lack of
high-energy emission does not automatically imply that the
reflection model is out of consideration. \citet{Piconcelli07}
indicated that the source does not display an iron line with a high
enough EW expected for a truly reflection-dominated spectrum.
Nonetheless, the reflection scenario a reflection spectrum does not
necessarily come along with a strong iron line, as it is not the
only feature produced by reflection. For instance, the Compton hump
and soft excess are signatures of reflection as well (though the
soft X-ray band of IRAS 09104+4109 is dominated by thermal emission
from the cluster gas). In a case that the iron abundance is not
high, the EW of the iron line can be low but the Compton hump is
present. Thus the EW of the iron line should not be considered as an
effective model selection tool.

\begin{figure}
\begin{center}
\leavevmode \epsfxsize=8.5cm \epsfbox{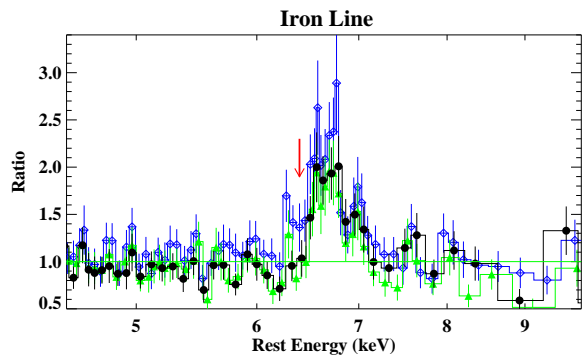}
\end{center}
\caption{The figure shows the data/model ratio of the
\emph{Chandra} 55-arcsec (blue open diamonds), the 2-55 arcsec (green
triangles), and the 5-55 arcsec (black points) spectra, which have been
fitted with a simple powerlaw. The red arrow points to the energy of
the 6.4 keV line.
              } \label{line_2}
\end{figure}

As mentioned in section \ref{spec_fitting}, Fig. \ref{model} clearly
shows that both models give similar fluxes over the $\sim$ 0.1-40.0
keV band. The reflection model produces a Compton hump which peaks
at $\sim$ 40 keV, while the transmission model predicts a steadily
rising spectral shape above 20 keV. This implies that
high-resolution X-ray spectroscopy in the hard-energy band may be
required to tell which model provides a better interpretation to the
data. If the spectrum is reflection-dominated, the shape of the
Compton hump should be detected. As the HXD/PIN failed to detect
emission above 10 keV, the only current space mission which is
likely to achieve this is \emph{NuSTAR}.

Another way to examine the possibility of each model is to
probe the origin of the neutral iron line. The 6.4 keV Fe K$\alpha$ line
could be fluorescent emission from (1) the accretion disc or (2)
nearby low-ionised gas or molecular clouds. If the iron line is
generated from disc reflection, it must come from the nucleus of the
source, where an AGN is accommodated. The line is likely reflected
from the edge of the cold outer disc as the torus. As for the latter possibility,
a cooling flow can form clouds of atomic or molecular gas. \citet{Churazov98} indicated
that the cold clouds illuminated by the X-ray emission of the hot
gas would lead to a 6.4 keV fluorescent line. \citet{O'Sullivan12}
show optical emission line filaments around the nucleus extended by a few arcsec,
indicating that cold gas clouds are present.

In order to examine the hypotheses mentioned above, we
generated \emph{Chandra} spectra that exclude different sizes of
central region and compared them with the \emph{Chandra} 55-arcsec
spectrum. We extracted a spectrum which excludes the central
2-arcsec-radius circular region, which means an extraction from a
2-55 arcsec annulus source region. Another spectrum excluding the
central 5-arcsec-radius (5-55 arcsec annulus source region) was
created using the same method. The background spectra were extracted
using a 55-110 arcsec annulus, the same as the background region we used
for the 55-arc spectrum extraction. We compare these spectra by
plotting the line profiles in Fig. \ref{line_2}, and the red arrow
indicates the 6.4 keV line in the rest frame. It can be seen that
there is a lack of strong emission at 6.4 keV rest frame in the 2-55
arcsec and the 5-55 arcsec spectra. We tested the transmission model
on these new spectra and found that the 6.4 keV Gaussian line is not
strongly required. The normalisation of the Gaussian component drops
to a number close to zero, and the model works equally well if the
Gaussian line is removed. This implies that most of the 6.4 keV
emission is contributed by the very centre of the source, that is,
the nucleus. The result cannot completely rule out the possibility
that the 6.4 keV line originates from the cold clouds, but shows
that the neutral iron line is more likely caused by the AGN. We
hereby consider the reflection scenario to be the better
interpretation to explain the X-ray spectrum of this source.

\section{Conclusion}

We carried out a \emph{Suzaku} observation which offers simultaneous
soft and hard X-ray band monitoring. During the observation we
avoided possible high-energy contamination from the nearby Type 2
AGN NGC 2785. Our result does not confirm the 3$\sigma$ detection by
the \emph{BeppoSAX} PDS instrument. By analysing both the latest
\emph{Suzaku} and \emph{Chandra} observations with long exposures,
we found that the broad iron line shown in these spectra is in fact
a combination of two components. Both the reflection and
transmission models have been tested on our data sets. Statistically
the data are consistent with either of the interpretations and both
models behave very similarly from the low-energy band to $\sim$ 40
keV. Nevertheless, both models indicate a Compton-thin absorber in
this source. Assuming that IRAS 09104+4109 does not change
significantly within a short time scale, it is a Compton-thin AGN.
Since the 6.4 keV Fe K$\alpha$ seems to come from the central AGN,
the reflection scenario is the better explanation to the source
spectrum. However, the X-ray data below 10 keV give degenerate
answers, and it is difficult to constrain the
flux above 10 keV by current observations. We need high spatial resolution X-ray spectroscopy such
as \emph{NuSTAR} to detect the Compton hump predicted by
the reflection model.

\section*{Acknowledgements}

CYC and ACF thank J. Hlavacek-Larrondo and K. Iwasawa for useful
discussions.

\bibliographystyle{mn2e_uw}
\bibliography{iras09104}

\label{lastpage}
\end{document}